\documentclass[aps,prl,twocolumn,groupedaddress,floatfix]{revtex4}

\usepackage[dvips]{graphicx,color} % Graphics and color
\usepackage{array,hhline,dcolumn} % Better table handling
\usepackage{rotating} % Rotate and sideways environments
\usepackage{verbatim}

\newcommand{\nubar}{\bar{\nu}}
\newcommand{\numu}{\nu_{\mu}}
\newcommand{\numubar}{\bar{\nu}_{\mu}}
\newcommand{\nue}{\nu_e}
\newcommand{\nuebar}{\bar{\nu}_e}
\newcommand{\nutau}{\nu_{\tau}}
\newcommand{\nutaubar}{\bar{\nu}_{\tau}}
\newcommand{\nuother}{\nu_{other}}

\newcommand{\gtwid}{\mathrel{\raise.3ex\hbox{$>$\kern-.75em\lower1ex\hbox{$\sim$}}}}
\newcommand{\ltwid}{\mathrel{\raise.3ex\hbox{$<$\kern-.75em\lower1ex\hbox{$\sim$}}}}

\begin{document}
% APS preprint designation
% \preprint{MiniBooNE_SN_Search}

\title{A Search for Core-Collapse Supernovae using the MiniBooNE Neutrino Detector}

\author{A.~A. Aguilar-Arevalo$^{13}$, C.~E.~Anderson$^{18}$,
        A.~O.~Bazarko$^{15}$, S.~J.~Brice$^{7}$, B.~C.~Brown$^{7}$,
        L.~Bugel$^{5}$, J.~Cao$^{14}$, L.~Coney$^{5}$,
        J.~M.~Conrad$^{12}$, D.~C.~Cox$^{9}$, A.~Curioni$^{18}$,
        Z.~Djurcic$^{5}$, D.~A.~Finley$^{7}$, M.~Fisher$^{8}$, B.~T.~Fleming$^{18}$,
        R.~Ford$^{7}$, F.~G.~Garcia$^{7}$,
        G.~T.~Garvey$^{10}$, J.~Grange$^{8}$, C.~Green$^{7,10}$, J.~A.~Green$^{9,10}$,
        T.~L.~Hart$^{4}$, E.~Hawker$^{3,10}$,
        R.~Imlay$^{11}$, R.~A. ~Johnson$^{3}$, G.~Karagiorgi$^{12}$,
        P.~Kasper$^{7}$, T.~Katori$^{9,12}$, T.~Kobilarcik$^{7}$,
        I.~Kourbanis$^{7}$, S.~Koutsoliotas$^{2}$, E.~M.~Laird$^{15}$,
        S.~K.~Linden$^{18}$,J.~M.~Link$^{17}$, Y.~Liu$^{14}$,
        Y.~Liu$^{1}$, W.~C.~Louis$^{10}$,
        K.~B.~M.~Mahn$^{5}$, W.~Marsh$^{7}$, C.~Mauger$^{10}$,
        V.~T.~McGary$^{12}$, G.~McGregor$^{10}$,
        W.~Metcalf$^{11}$, P.~D.~Meyers$^{15}$,
        F.~Mills$^{7}$, G.~B.~Mills$^{10}$,
        J.~Monroe$^{5}$, C.~D.~Moore$^{7}$, J.~Mousseau$^{8}$, R.~H.~Nelson$^{4}$,
        P.~Nienaber$^{16}$, J.~A.~Nowak$^{11}$,
        B.~Osmanov$^{8}$, S.~Ouedraogo$^{11}$, R.~B.~Patterson$^{15}$,
        Z.~Pavlovic$^{10}$, D.~Perevalov$^{1}$, C.~C.~Polly$^{9,7}$, E.~Prebys$^{7}$,
        J.~L.~Raaf$^{3}$, H.~Ray$^{8,10}$, B.~P.~Roe$^{14}$,
        A.~D.~Russell$^{7}$, V.~Sandberg$^{10}$, R.~Schirato$^{10}$,
        D.~Schmitz$^{5}$, M.~H.~Shaevitz$^{5}$, F.~C.~Shoemaker$^{15}$\footnote{deceased},
        D.~Smith$^{6}$, M.~Soderberg$^{18}$,
        M.~Sorel$^{5}$\footnote{Present address: IFIC, Universidad de Valencia and CSIC, Valencia 46071, Spain},
        P.~Spentzouris$^{7}$, J.~Spitz$^{18}$, I.~Stancu$^{1}$,
        R.~J.~Stefanski$^{7}$, M.~Sung$^{11}$, H.~A.~Tanaka$^{15}$,
        R.~Tayloe$^{9}$, M.~Tzanov$^{4}$,
        R.~G.~Van~de~Water$^{10}$, 
        M.~O.~Wascko$^{11}$\footnote{Present address: Imperial College; London SW7 2AZ, United Kingdom},
         D.~H.~White$^{10}$,
        M.~J.~Wilking$^{4}$, H.~J.~Yang$^{14}$,
        G.~P.~Zeller$^{5,10}$, E.~D.~Zimmerman$^{4}$ \\
\smallskip
(The MiniBooNE Collaboration)
\smallskip
}
\smallskip
\smallskip
\affiliation{
$^1$University of Alabama; Tuscaloosa, AL 35487 \\
$^2$Bucknell University; Lewisburg, PA 17837 \\
$^3$University of Cincinnati; Cincinnati, OH 45221\\
$^4$University of Colorado; Boulder, CO 80309 \\
$^5$Columbia University; New York, NY 10027 \\
$^6$Embry Riddle Aeronautical University; Prescott, AZ 86301 \\
$^7$Fermi National Accelerator Laboratory; Batavia, IL 60510 \\
$^8$University of Florida; Gainesville, FL 32611 \\
$^9$Indiana University; Bloomington, IN 47405 \\
$^{10}$Los Alamos National Laboratory; Los Alamos, NM 87545 \\
$^{11}$Louisiana State University; Baton Rouge, LA 70803 \\
$^{12}$Massachusetts Institute of Technology; Cambridge, MA 02139 \\
$^{13}$Instituto de Ciencias Nucleares, Universidad National Aut\'onoma de M\'exico, D.F. 04510, M\'exico \\
$^{14}$University of Michigan; Ann Arbor, MI 48109 \\
$^{15}$Princeton University; Princeton, NJ 08544 \\
$^{16}$Saint Mary's University of Minnesota; Winona, MN 55987 \\
$^{17}$Virginia Polytechnic Institute \& State University; Blacksburg, VA
24061
\\
$^{18}$Yale University; New Haven, CT 06520\\
}

\date{\today}

\begin{abstract}

We present a search for core-collapse supernovae in the Milky Way galaxy, 
using the MiniBooNE neutrino detector.  No evidence is found for core-collapse 
supernovae 
occurring in our Galaxy in the period from December 14, 2004 to July 31, 2008, corresponding to 
98\% live-time for collection.  We set a limit on the 
core-collapse supernova rate out to a distance of 13.5 kpc to be less than 0.69 supernovae per 
year at 90\% CL.  
\end{abstract}


% Use showkeys class option if keyword
% display desired
\maketitle

%%%%%%%%%%%%%%%%%%%%%%%%%%%%%%%%%%%%%%%%%%%%%%%%%%%%%%%%%%%%%%%%%%%%%%%%%%%%%%%%%
%%%%%%%%%%%%%%%%%%%%%%%%%%%%%%%%%%%%%%%%%%%%%%%%%%%%%%%%%%%%%%%%%%%%%%%%%%%%%%%%%
\section{Introduction}

Supernovae are stars that explode and become extremely luminous.  Core-collapse 
supernovae typically begin as stars with masses greater than 8M$_{\odot}$.  
When these stars explode, $\sim 3 \times 10^{53}$ ergs of gravitational binding energy is 
released in a burst of neutrinos and anti-neutrinos lasting approximately 
10 seconds~\cite{Masayuki}.  This neutrino burst will arrive at the Earth several hours prior to photons from 
the supernova; neutrino detectors can be utilized as
 the first line of detection for supernovae.

Although current predictions for the rate of supernovae in the Milky Way galaxy
 are between 1 and 12 per century~\cite{GoodBook}, 
the last observed supernova in our Galaxy occurred in 1604, in the constellation 
Ophiuchus~\cite{kepler}.  Detection of supernovae using optical telescopes is 
highly dependent on the orientation of the telescope with respect to the supernova.  Neutrino detectors are 
able to observe supernovae occurring at any point in our Galaxy, 
regardless of the orientation of the supernova with respect to the detector.
Starting in the late 1990's, a network composed of neutrino detectors has been performing a real-time search
for supernovae~\cite{snews}.  When a supernova is observed this network will provide coordinates to 
observatories across the world, allowing them to align their telescopes in time to observe the photons 
from the supernova.

Supernova neutrino data can be used to verify astronomical predictions of the stellar collapse 
model and to provide bounds on standard model quantities.  Detection of the neutrinos 
from SN1987A in the Large Magellanic Cloud by the 
Kamiokande-II~\cite{k2k} and IMB~\cite{IMB} water Cerenkov experiments provided 
upper limits on the lifetime and mass of the $\nuebar$ that were
 comparable to results from laboratory-based experiments at the time~\cite{Masayuki}.  

Several neutrino detectors have published results from their 
search for supernovae occurring in our Galaxy~\cite{othersearches}.  The LVD detector, 
located at the Gran Sasso Underground Laboratory in Italy, set an 
upper limit of 0.18 supernovae per year in our Galaxy at 90\% C.L., 
for 14 years of run-time, from 1992 to 2006~\cite{LVD}.  
The Super-Kamiokande experiment in Japan 
set a limit of less than 0.32 supernovae per year at 
a distance of 100 kpc, at the 90\% CL for the period of time of 
May, 1996 to July, 2001 and December, 2002 to 
October, 2005~\cite{superk}.  MiniBooNE's search covers a more recent period of time, 
from December, 2004 to July, 2008.  

%%%%%%%%%%%%%%%%%%%%%%%%%%%%%%%%%%%%%%%%%%%%%%%%%%%%%%%%%%%%%%%%%%%%%%%%%%%%%%%%%
%%%%%%%%%%%%%%%%%%%%%%%%%%%%%%%%%%%%%%%%%%%%%%%%%%%%%%%%%%%%%%%%%%%%%%%%%%%%%%%%%
\section{The MiniBooNE Experiment}

	MiniBooNE is a neutrino experiment designed to search for $\numu \rightarrow \nue$ 
and $\numubar \rightarrow \nuebar$ neutrino oscillations, using a beam of neutrinos produced 
by the Booster beam-line at Fermi National Accelerator 
Laboratory~\cite{flux}.  
The MiniBooNE detector is a spherical tank of inner radius 610 cm, filled with 800 tons of 
mineral oil (CH$_2$)~\cite{MB}.  An optical barrier divides the detector into two regions.  The inner region
 contains 1280 inward-facing photomultiplier tubes (PMTs), providing 
10\% photocathode coverage.  The outer region is lined with 240 PMTs that provide 
a veto for charged particles entering or leaving the tank, such as cosmic rays~\cite{QE}.  
The detector is buried 3 meters underground, 
at a sufficient distance to eliminate the majority of incoming cosmic ray hadrons.  
However, a 10 kHz rate of cosmic ray muons can still penetrate this barrier, and their progeny 
are the main source of 
background for the supernova search.

%%%%%%%%%%%%%%%%%%%%%%%%%%%%%%%%%%%%%%%%%%%%%%%%%%%%%%%%%%%%%%%%%%%%%%%%%%%%%%%%%
%%%%%%%%%%%%%%%%%%%%%%%%%%%%%%%%%%%%%%%%%%%%%%%%%%%%%%%%%%%%%%%%%%%%%%%%%%%%%%%%%
\section{Theoretical Predictions}

The prediction for the observation of a supernova signal in MiniBooNE is based on the following 
assumptions~\cite{Sharp}:

\begin{itemize}
\item The event is a core-collapse supernova.
\item When the core collapses and rebounds, the change in the gravitational binding energy 
is $\sim 3\times10^{53}$ ergs.
\item Each of the 6 types of neutrinos ($\nue, \nuebar, \numu, \numubar, \nutau, \nutaubar$) 
carry away 1/6 of this binding energy.
\item The neutrinos are emitted during a 10 second burst.
\item The neutrinos are emitted isotropically.
\item The neutrinos produced during a supernova core collapse have the 
following average energies and temperatures~\cite{GoodBook}: 
$\langle\mathrm{E}_{\nu_e}\rangle \approx $ 10 MeV and $T \approx$ 3.5 MeV,   
$\langle\mathrm{E}_{\bar{\nu}_e}\rangle \approx$ 15 MeV and $T \approx$ 5 MeV, 
and $ \langle\mathrm{E}_{\nu_{other}}\rangle \approx$ 20 MeV and $T \approx$ 8 MeV, 
where other is $\nu_\mu, \bar{\nu}_{\mu}, \nu_\tau$, and $\bar{\nu}_{\tau}$.
\item The spectrum for supernova events is characterized by a Fermi-Dirac distribution with zero chemical potential.
\end{itemize}

The charged current (CC) interaction, $\nubar + p \rightarrow l^{+} + n$ and $\nu + n \rightarrow l^{-} + p$, 
is used to search for the supernova signal in MiniBooNE.  The $\nuother$ aren't energetic enough to engage in 
this interaction.  The $\nuebar$ 
reaction will dominate the event sample in our detector, due to the larger cross section 
of this interaction on free protons present in the CH$_2$ molecules in 
mineral oil.  Therefore, higher-energy $\nuebar$s 
are the primary constituent of the detectable supernova neutrino signal in MiniBooNE.  

The predicted supernova rate in MiniBooNE assumes a maximum usable detector radius of 550 cm, 25 cm inside the optical barrier.  Event locations are reconstructed from light intensity and timing information from the PMT array.

\subsection{Predicted Signal}

The number of expected signal events is described by:

\begin{equation}
\footnotesize
N = 11.8\left(\frac{E_B}{10^{53}\mathrm{erg}}\right)\left(\frac{\mathrm{1 MeV}}{T}\right)\left(\frac{ \mathrm{10 kpc}}{D}\right)^2\left(\frac{M_D}{\mathrm{1 kton}}\right)\left(\frac{\langle\sigma\rangle}{10^{-42}\mathrm{cm}^2}\right),
\label{eqn}
\end{equation}

\noindent
where $E_B$ is the binding energy released during the supernova, $T$ is the temperature of the emitted $\nuebar$, D is the distance to the supernova, $M_D$ is the fiducial mass of the detector, and $\langle\sigma\rangle$ is the thermally averaged free proton cross section for $\nuebar + p \rightarrow e^{+} + n$.

\begin{table*}
\begin{tabular}{|c||c|c|}
\hline
\textbf{\em Symbol}  &  Meaning   &   Value  \\
\hline \hline
$E_B$ & Gravitational binding energy of Supernova & $3 \times 10^{53} \mathrm{ergs}$ \\
$T$  & Temperature of incoming neutrinos & 5 MeV  \\
$D$  & Distance of core-collapse supernova from Earth & 10 kpc   \\
$\langle\sigma\rangle$ & Thermally-averaged cross section in mineral oil & $54 \times 10^{-42} \mathrm{cm}^2$ \\
$M_D$ & Fiducial mass of the detector & 0.595 ktons @ r = 5.5 m \\
$M_D$, final event selection     &        & 0.326 ktons @ r = 4.5 m   \\
\hline
\end{tabular}
\caption{Symbol information for Equation~\ref{eqn}}
\label{tab:t}
\end{table*}

The two parameters, $T$ and $E_B$, have uncertainties associated 
with them.  Observational limits set by SN1987A constrain the amount 
that one can vary these parameters individually and simultaneously.  
For individual variations, $E_B$ varies between 2 $\times$ 10$^{53}$ and 3 $\times$ 10$^{53}$ 
ergs, and the temperature 
lies between 4 and 6 MeV~\cite{beacom}.  A discussion of the systematic error assigned to 
our result due to these parameters is presented later.  

Using the parameter values listed in Table~\ref{tab:t}, 226 of these events are expected 
in MiniBooNE with a reconstructed lepton energy of 0-60 MeV, using a fiducial detector radius of 550 cm 
and prior to any event selection cuts~\cite{Sharp}.

The neutral current interaction, $^{12}$C($\nu, \ \nu^{\prime}$)$^{12}$C$^*$ (15.11 MeV), 
produces a 
15.11 MeV photon that appears electron-like in our detector.  This interaction may increase our 
event rate by $\sim$23 events~\cite{borexino}.  
However, this number is less than the 
uncertainty on the prediction due to the temperature~\cite{beacom2}, and thus this interaction 
channel is not included in this analysis.

By assuming neutrino emission parameters consistent with SN1987A data, the predicted 
number of signal events includes effects due to neutrino mixing prior to arrival 
at the MiniBooNE detector.

\subsection{Predicted Sources of Background Events}

The decay products of cosmic rays are predicted to be the only source of 
background to the supernova signal~\cite{Sharp}.  
They produce two distinct backgrounds to this search: Michel electrons and $^{12}$B.  
Stopped cosmic ray muons occur in the detector at a rate of 2 kHz.  
Approximately 95$\%$ of these stopped muons will decay to Michel electrons or positrons whose
 energy spectrum has an endpoint of 52.8 MeV.  With trigger-level cuts applied, 
the rate of these background events is reduced to 2 Hz.  

Forty-four percent of the 2 kHz of stopped cosmic ray muons are $\mu^{-}$.  Eight percent of these 
$\mu^{-}$ will capture on $^{12}$C nuclei in the detector's mineral oil, 
of which 16$\%$ will become particle unbound states of $^{12}$B.  
This isotope of boron is unstable to $\beta^{-}$ decay.  
The electrons produced in this interaction occur with a frequency of 11 Hz 
(=2 kHz $\times$ 0.44 $\times$ 0.08 $\times$ 0.16), 
and have an energy spectrum with an endpoint of 13.9 MeV.

The MiniBooNE supernova search avoids the need for any detailed knowledge of the backgrounds, 
as further explained in the Results section.

%%%%%%%%%%%%%%%%%%%%%%%%%%%%%%%%%%%%%%%%%%%%%%%%%%%%%%%%%%%%%%%%%%%%%%%%%%%%%%%%%\
%%%%%%%%%%%%%%%%%%%%%%%%%%%%%%%%%%%%%%%%%%%%%%%%%%%%%%%%%%%%%%%%%%%%%%%%%%%%%%%%%
\section{Analysis Details}

The supernova search was performed on data spanning the period from 12/14/2004 to 07/31/2008.  
The data are broken down into runs that typically last a few hours.  Runs are composed 
of events, each lasting 19.6 microseconds and triggered by a particular set of conditions being met 
by the PMT and external signals.  
It is prohibitively time consuming to apply the complete set of event selection cuts 
to every 10 second window of events in this data 
sample.  This problem is circumvented by performing two passes over the data.  The first pass over the data 
applies low level cuts that remove time windows with low event counts.  
The second pass applies quality cuts based on reconstructed quantities to isolate the predicted 
supernova signal from expected background sources.

\subsection{Event Selection Cuts: First Pass}

The first pass over the data must meet two requirements: the beam-off activity trigger, and a data quality 
filter.  The beam-off activity trigger separates potential supernova neutrino events 
from events occurring in the detector from other sources.  It is set when the following conditions 
are met: 
\begin{itemize}
\item Time since last neutrino beam event $>$ 20 $\mu$s.
\item Number of inner detector PMT hits $\geq$ 60.
\item Number of veto region PMT hits $<$ 6.
\item The time since the number of tank hits is $\geq$ 100 is $\geq$ 15 $\mu$s, 
and the time since the last activity in the veto region is $\geq$ 15 $\mu$s.
\end{itemize}

The first condition functions as a filter to remove neutrino beam data.  
The second condition serves as a lower bound on the detected energy. 
The third and fourth conditions reject cosmic ray muon events and temporarily disable data recording 
for 15 $\mu$s, enough time for a muon ($\tau_\mu \approx 2.2 \ \mu$s) 
to decay and its decay products to cease interacting in the detector.
The data quality filter reduces our live time by 2\%.

The total data set consists of 6997 runs.  
After selection, the data are further split into 10 second intervals.  These 
are sliding intervals; each event within a run starts a new 10 second window.  
The number of events per 10 second window is recorded and histogrammed.  
Let $\mu$ and $\sigma$ denote the mean and the standard deviation of this histogram, respectively.  Any run 
containing a 10 second time period with more than Z events, where Z = $\mu$ + 5$\sigma$, 
is identified as 
containing a potential supernova candidate and selected to continue to the second pass.

\begin{figure}[h]
\scalebox{0.35}{\includegraphics[angle=-90]{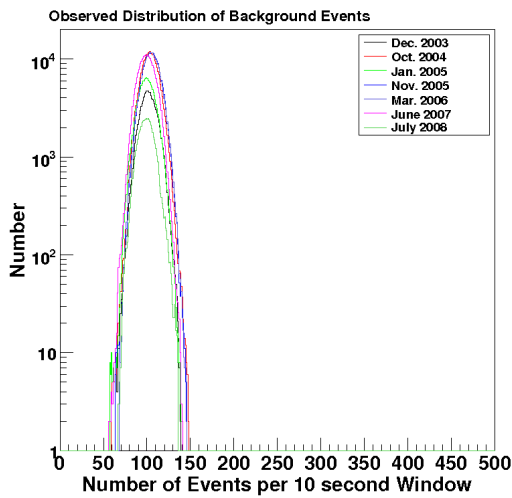}}
    \caption{Number of background events per 10 second window passing the beam-off activity trigger and the data quality filter for runs spanning our entire data sample, broken into time intervals.  Runs shown were not selected during the first pass.  Each shape contains $\approx$10$^5$ 10 second windows.}
    \label{fig:backgaus}
\end{figure}

The examination of runs not selected provides a measure of the background distribution of events.  
The background distribution of events at this stage of the analysis has a 
mean of $\mu$$\approx$100 and a standard deviation of $\sigma$$\approx$10 events, per 10 second window 
(See Figure~\ref{fig:backgaus}).  
These numbers have remained stable throughout the MiniBooNE data collection period for all runs, 
rejected and selected, in the beam-off activity sample.  
This first pass selects time windows containing greater than Z=150 events per 10 second window.  
This serves to greatly reduce the dataset, in preparation for the next step of the analysis.

\subsection{Event Selection Cuts: Second Pass}
After potential supernova candidates are identified using the first pass over the data, 
event selection cuts are used to isolate the potential supernova signal from the background events. 

The number of bursts of light, separated by time within an event, are referred to as sub-events.  More than one sub-event is indicative of multiple interaction products.  Only one burst of light will be created in the detector from positrons  
produced by the $\nuebar$ CC interaction.  The number of veto PMT hits must be less than 6, to 
remove potential cosmic ray events.  
The number of tank hits is roughly proportional to the energy of the particles interacting 
inside the detector.  The possible range of tank hits for a supernova signal is 50 to 200.  
Any number above 200 is more indicative of a $\mu$ in the detector, and any amount lower than 50 
is indicative of a low energy background event.

Energy cuts based on the expected Michel electron and $^{12}$B backgrounds 
are applied to further reduce the number of background events.  
An energy cut of 11-45 MeV allows for minimal loss of signal 
while maintaining a fairly large signal to background ratio.  
Finally, the neutrino interaction must take place in the inner 450 cm of our detector.
This more restrictive fiducial requirement removes low-energy background events that penetrate from the outside 
of the detector.

A summary of the cuts applied during the second pass is:
\begin{itemize}
\item Beam-off activity trigger. 
\item Data quality filter.
\item 1 burst of light (subevent). 
\item Number of veto region PMT hits $<$ 6.
\item 50 $<$ Number of inner region PMT hits $<$ 200.
\item Reconstructed radius $<$ 450 cm.
\item 11 MeV $<$ Reconstructed lepton energy $<$ 45 MeV.  
\end{itemize}

The effect of these cuts is to reduce the expected number of supernova events from 226 to 110 
for a supernova at a distance of 10 kpc.

After these cuts are applied, the data are split into 10 second intervals, 
and the same selection procedure as applied in the first pass is repeated.  The examination 
of runs that are not selected 
indicates the background distribution at this stage of the analysis has a 
mean of $\mu$$\approx$20 and a standard deviation of $\sigma$$\approx$4.  

%%%%%%%%%%%%%%%%%%%%%%%%%%%%%%%%%%%%%%%%%%%%%%%%%%%%%%%%%%%%%%%%%%%%%%%%%%%%%%%%%
%%%%%%%%%%%%%%%%%%%%%%%%%%%%%%%%%%%%%%%%%%%%%%%%%%%%%%%%%%%%%%%%%%%%%%%%%%%%%%%%%
\section{Results}

During the first pass through the data, we identify 319 out of 6997 run numbers 
as containing potential supernova candidates.  
The data from these runs are processed using the full set of cuts, and 78 of the 319 
runs remain.  

\begin{figure}[h]
\scalebox{0.35}{\includegraphics[angle=-90]{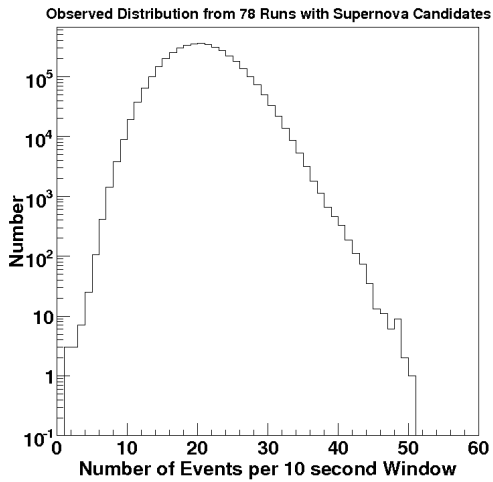}}
    \caption{Distribution of the number of neutrino events per 10 second window, for all 78 runs identified as containing potential supernova candidates.  There are no windows with greater than 51 total events per window.}
    \label{fig:sum}
\end{figure}

The distribution of the number of events per 10 second window for all 78 remaining runs 
is shown in Figure~\ref{fig:sum}, with a mean of 20.11 events and a standard 
deviation of 4.43 events.  There are no cases where $>$51 events per 10 second window are observed.

The background event estimate is determined by examining time windows in the beam-off 
trigger stream that pass the data quality filter, but 
that are not selected during the first pass.  The entire set of event 
selection cuts are applied 
to a set of runs spanning the entire collection period, from December, 
2004 through July, 2008.  The mean of the distribution is 20.34.  
Therefore, we set a limit for supernova observation based on (52 - 20.34), or 32 signal events.

One hundred and ten signal events are expected from a supernova occurring within 10 kpc of MiniBooNE, 
after applying all event selection cuts.  
The null result is used to place a limit on the probability that a 
supernova occurred during our search window.  This approach, though conservative, has the benefit of 
not requiring a prediction for the background energy distribution.
The null observation of 32 events allows us to set a limit for a distance 
greater than 10 kpc.  Equation~\ref{eqn}, which describes the relation between expected events and 
distance from the supernova, is adjusted to account for efficiency of the event selection cuts.  
The number of expected events represents the mean of a Gaussian, with a root-mean-square of 
$\sqrt{\mathrm{N}}$, or the statistical error on N.  The uncertainty on this number is 
driven by uncertainties on the two parameters, $T$ and $E_B$.  
The free proton cross section is proportional to $T$$^2$, 
making Equation~\ref{eqn} proportional to $T$ $\times$ $E_B$.  Simple error propagation results in a systematic 
error of 0.36N.  However, the observation of SN1987A constrains 
the amount that $T$ and $E_B$ can fluctuate simultaneously.  
Consequently, the total systematic error assigned is 0.26N.
  
\begin{figure}[t]
\scalebox{0.35}{\includegraphics[angle=-90]{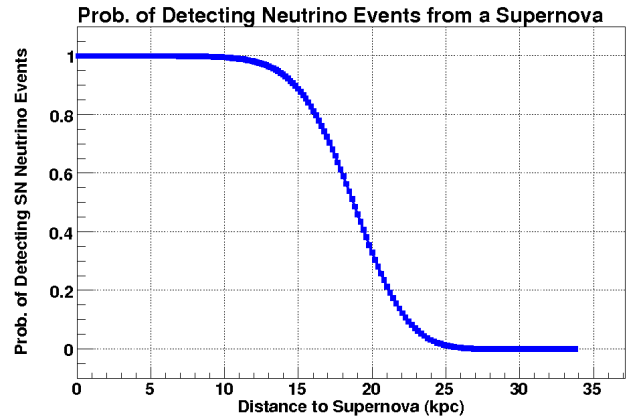}}
    \caption{Probability of detecting a supernova as a function of distance in kpc.  A fiducial radius of 450 cm and the efficiency of the event selection cuts are included.  Statistical and systematic errors are included.}
    \label{fig:prob}
\end{figure}

The detection probability was formed by calculating the probability that the
 Gaussian with a mean of N and root-mean-square of $\sqrt{N + (0.26N)^2}$ could fluctuate down to 32.  
Figure~\ref{fig:prob} shows the calculated detection 
probability, as a function of distance to the supernova.  The limit is set at the point 
where our detection efficiency drops to 95\%.  
Our limit of 32 signal events corresponds to a distance of 13.5 kpc.  This limit increases to 
16.2 kpc in the absence of systematic errors.

Using a Poisson probability distribution, the observation of 0 events over one collection period 
allows us to set a limit of 2.3 supernovae at the 90\% CL.  Following the example of 
the Super-Kamiokande search~\cite{superk}, we set a limit using the number of total live days in 
our data sample.  
The data sample used in this analysis corresponds to 1221.44 live days.  
Therefore, we set a limit at the 90\% CL on a supernova having occurred within 13.5 kpc of our detector, 
at a rate of 0.69 supernovae per year.  This limit corresponds to 73.8\% coverage of the Milky Way~\cite{miri}.

%%%%%%%%%%%%%%%%%%%%%%%%%%%%%%%%%%%%%%%%%%%%%%%%%%%%%%%%%%%%%%%%%%%%%%%%%%%%%%%%%
%%%%%%%%%%%%%%%%%%%%%%%%%%%%%%%%%%%%%%%%%%%%%%%%%%%%%%%%%%%%%%%%%%%%%%%%%%%%%%%%%
\section{Conclusions}

The search for supernovae using neutrino detectors is complementary, and in many ways superior, 
to searches performed using telescopes.  Using the MiniBooNE detector, 
we performed a search for supernovae using 
data taken between the period from 12/14/2004 to 07/31/2008.  A limit is set on the rate of 
core-collapse supernovae 
in the Milky Way within a distance of 13.5 kpc from the Earth to be less than 0.69 supernova per 
year at the 90\% CL.  This limit corresponds to 73.8\% coverage of the Milky Way~\cite{miri}.

%%%%%%%%%%%%%%%%%%%%%%%%%%%%%%%%%%%%%%%%%%%%%%%%%%%%%%%%%%%%%%%%%%%%%
%\newpage
%\clearpage

\begin{acknowledgments}
We acknowledge the support of Fermilab, the Department of Energy,
and the National Science Foundation. We are grateful to John Beacom for his valuable insight and advice.  
We thank Alessandro Mirizzi, Georg G. Raffelt, and Pasquale D. Serpico for providing the probability 
distribution for the Milky Way.
\end{acknowledgments}

\bibliography{prl}% Produces the bibliography via BibTeX.

\begin{thebibliography}{10}

\bibitem{Masayuki}
K. Zuber, \underline{Neutrino Physics}, Institute of Physics, London, 2004.

\bibitem{GoodBook}
C. Giunti and C. W. Kim, \underline{Fundamentals of Neutrino Physics and Astrophysics}, 
Oxford Press, New York, 2007.

\bibitem{kepler}
http://seds.lpl.arizona.edu/messier/more/mw\_sn.html.


\bibitem{snews}
K. Scholberg, 	
Astron. Nachr. 329, 337-339 (2008), 
arXiv:astro-ph/0803.0531v1.


\bibitem{k2k}
K. S. Hirata {\it et al}. [Kamiokande Collaboration], 
Phys. Rev. D 38, 448-458 (1988).


\bibitem{IMB}
R. M. Bionta {\it et al}. [IMB Collaboration], 
Phys. Rev. Lett. 58, 1494 (1987).

\bibitem{othersearches}
M. L. Cherry {\it et al.}, J. Phys. G 8, 879 (1982); 
G. T. Zatsepin and O. G. Ryazhskaya, Usp. Fiz. Nauk 146, 713 (1985) [Sov. Phys. Usp. 28, 726 (1985)]; 
E. N. Alexeyev {\it et al.}, Zh. Eksp. Teor. Fiz. 104, 2897 (1993) [JETP 77, 339 (1993)]; 
R. S. Miller {\it et al.}, Astrophys. J. 428, 629 (1994);
M. Ambrosio {\it et al.}, Astropart. Phys. 8, 123 (1998);
R. V. Novoseltseva {\it et al.}, arXiv:0910.0738v1 [astro-ph.HE].

\bibitem{LVD}
M. Selvi {\it et al}. [LVD Collaboration], 
arXiv:hep-ex/0608061v1.

\bibitem{superk}
M. Ikeda {\it et al}. [Super-Kamiokande Collaboration],
Astrophys. J. 669, 519-524 (2007), 
arXiv:astro-ph/0706.2283v1.


\bibitem{flux}
A.A. Aguilar-Arevalo {\it et al}. [MiniBooNE Collaboration],
Phys. Rev. D 79, 072002 (2009),
arXiv:hep-ex/0805.1764.


\bibitem{MB}
A.A. Aguilar-Arevalo {\it et al}. [MiniBooNE Collaboration],
Nucl. Instr. Meth. A599, 28-46 (2009),
arXiv:hep-ex/0806.4201.


\bibitem{QE}
A.A. Aguilar-Arevalo {\it et al}. [MiniBooNE Collaboration],
Phys. Rev. Lett. 100, 032301 (2008), 
arXiv:hep-ex/0706.0926.


\bibitem{Sharp}
M. Sharp, J. Beacom, J. Formaggio,
Phys. Rev. D 66, 013012 (2002),
arXiv:hep-ph/0205035v2.


\bibitem{beacom} W. D. Arnett {\it  et al}.,
Ann. Rev. Astron. Astrophys. 27, 629-700 (1989). 


\bibitem{borexino} L. Cadonati, F. P. Calaprice, M. C. Chen, 
Astropart. Phys. 16, 361-372 (2002), 
arXiv:hep-ph/0012082v1. 


\bibitem{beacom2} J. Beacom, W. Farr, P. Vogel, 
Phys. Rev. D 66, 033001 (2002),
arXiv:hep-ph/0205220v1.


\bibitem{miri} A. Mirizzi, G. G. Raffelt, P. D. Serpico,
arXiv:astro-ph/0604300v2.

\end{thebibliography}

\end{document}